\documentclass[twocolumn,showpacs,prl]{revtex4}
\usepackage{times,xspace}
\usepackage{amsbsy,amssymb,amsmath,bm}
\usepackage{graphicx,color,epsfig,rotate}
\usepackage{fancyhdr}

\def\bbbc{{\mathchoice {\setbox0=\hbox{$\displaystyle\rm C$}\hbox{\hbox
to0pt{\kern0.4\wd0\vrule height0.9\ht0\hss}\box0}}
{\setbox0=\hbox{$\textstyle\rm C$}\hbox{\hbox
to0pt{\kern0.4\wd0\vrule height0.9\ht0\hss}\box0}}
{\setbox0=\hbox{$\scriptstyle\rm C$}\hbox{\hbox
to0pt{\kern0.4\wd0\vrule height0.9\ht0\hss}\box0}}
{\setbox0=\hbox{$\scriptscriptstyle\rm C$}\hbox{\hbox
to0pt{\kern0.4\wd0\vrule height0.9\ht0\hss}\box0}}}}

\newcommand{\ignore}[1]{}
\newcommand{\mComment}[1]{}
\newcommand{\gComment}[1]{}
\newcommand{\jComment}[1]{}
\newcommand{\rComment}[1]{}
\newcommand{\lComment}[1]{}

\renewcommand{\mComment}[1]{\textcolor{blue}{Bruce: #1}}
\renewcommand{\gComment}[1]{\textcolor{red}{Zohar: #1}}
\renewcommand{\jComment}[1]{\textcolor{green}{Cristian: #1}}

\begin{document}
\title{Effective Low-Energy Model for $f$-electron Delocalization}
\author{K. A. Al-Hassanieh, Yi-feng Yang, Ivar Martin and C. D. Batista}
\affiliation{Los Alamos National Laboratory, Los Alamos, New Mexico 87545}
\date{\today}

\begin{abstract}
We consider a Periodic Anderson Model (PAM) with a momentum-dependent inter-band hybridization 
that is strongly suppressed near the Fermi level. Under these conditions, we reduce the PAM to an effective low-energy
Hamiltonian, $H_{\rm eff}$, by expanding in the small parameter $V_0/t$ ( $V_0$ is the maximum inter-band hybridization amplitude and $t$ is the hopping integral of the broad band). The resulting model consists of  a $t-J$ $f$-band coupled 
via the Kondo exchange to the electrons in the broad band. $H_{\rm eff}$ allows for studying the $f$-electron delocalization transition. 
The result is a doping-induced Mott transition for the $f$-electron delocalization, 
which we demonstrate by density-matrix renormalization group (DMRG) calculations.
\end{abstract}

\pacs{72.80.Sk, 74.25.Ha, 73.22.Gk}
\maketitle


A recent de Haas-van Alphen (dHvA) experiment on CeIn$_3$ revealed small $f$-character hole-pockets  \cite{Sebastian09} that coexist
with local-moment antiferromagnetism (AFM)  \cite{Lawrence80}. 
This observation defies the conventional view of the heavy fermion materials according to which large-moment magnetism exists only for strongly localized f-electrons\cite{Doniach77}.  Therefore, a new paradigm for the interplay between f- and conduction electrons needs to be developed.  

The Periodic Anderson model (PAM) is the minimal Hamiltonian for describing actinide and lanthanide based compounds. 
The model includes a periodic array of strongly interacting $f$-orbitals coupled to a broad conduction band via a hybridization amplitude $V_{\bf k}$.  Inter-band charge fluctuations are significantly suppressed when the bare $f$-electron level, $\epsilon_f$, is well below the Fermi energy of the conduction band. In this localized regime, the PAM can be reduced to a Kondo lattice model (KLM) by means of a Schrieffer-Wolff transformation \cite{Schrieffer66}. The $f$-electrons act like local magnetic moments that can either order, or ``dissolve'' into the Fermi sea of the conduction band, leading to a large increase of the quasi-particles effective mass (heavy fermion phase) \cite{Fisk88}. When $\epsilon_f$ gets closer to the Fermi level, $\mu$, the charge transfer between the bands increases and finally leads to $f$-electron delocalization. In general, the Schrieffer-Wolff transformation cannot be extended to this {\it mixed-valence regime}, when  $|V_{\bf k}|$ becomes comparable to $|\epsilon_f-\mu|$. 
Consequently, the lack of a control parameter poses a challenge for describing the crossover between the localized and mixed-valence regimes.

Here we consider a particular case  that allows for extending the Schrieffer-Wolff transformation to the mixed-valence regime. 
The basic assumption is that $V_{\bf k}$ cancels at the crossing points 
between the two bands. Under this assumption, we derive under control a low-energy effective model that extends and generalizes the KLM. It consists of a $t-J$ model in the $f$-band  coupled via Kondo exchange to the broad-band electrons. The new model describes the continuous crossover between the localized and mixed valence regimes. We show that there are two general classes of the low-energy behavior of the PAM, depending on whether the inter-band or  intra-$f$-band coherent charge fluctuations dominate. The former -- ``Kondo" -- regime is characterized by strong hybridization between the conduction and the $f$ bands and can lead either to AFM or heavy fermi liquid ground state \cite{Doniach77}. In the latter -- ``Mott" -- regime, the {\it $f$-electron delocalization is very similar to the doping-induced Mott transition}. The role of the conduction band is primarily limited to providing a charge reservoir for the correlated $f$-band. Consequently, the $f$-electron delocalization is accompanied by a  
change in the Fermi surface topology (Lifshitz transition). Unlike the standard Kondo regime, new $f$-character pockets emerge at the transition and coexist with the large Fermi surface of the broad band.  According to calculations which are controlled only in the zero doping limit \cite{Trugman90,Dagotto94}, these $f$-character pockets seem to be the characteristic Fermi surface of lightly doped Mott insulators (MIs). The heaviness of the $f$-pockets is caused by magnetic frustration of the kinetic energy \cite{Trugman90}. This scenario provides a possible explanation for the recent dHvA in CeIn$_3$ \cite{Sebastian09}.

We start by considering a PAM Hamiltonian of the form:
$
H=H_0+H_I,
$
with
\begin{eqnarray}
H_0&=& -t\sum_{\langle j,l\rangle\sigma} (c^{\dagger}_{j\sigma}c^{\;}_{l\sigma}+ \text{H.c.}) -\mu\sum_{j\sigma} n_{j\sigma} 
\nonumber \\
&+& \sum_{j\sigma} (\epsilon_f - \mu) n^f_{j\sigma} + U\sum_{j} n^f_{j\uparrow}n^f_{j\downarrow},
\nonumber \\
H_I&=& \sum_{ j,l} V_{jl} (c^\dagger_{j\sigma} f^{\;}_{l\sigma} + \text{H.c.}),
\label{pam}
\end{eqnarray}
where $\langle j , l \rangle$ indicates that $j$ and $l$ are nearest-neighbor sites,
$f^{\dagger}_{j\sigma}$ ($c^\dagger_{j \sigma}$) creates an $f$ ($c$)-electron with
spin $\sigma$ on site $j$, $n^f_{j\sigma} = f^{\dagger}_{j\sigma} f_{j\sigma}$, 
and $n_{j\sigma} = c^{\dagger}_{j\sigma}c^{\;}_{j\sigma}$.
The chemical potential $\mu$ controls the total number of electrons.
The $f$-electrons interact via an on-site Coulomb repulsion $U$.
$H_0$ contains the terms that do not mix the two bands, while $H_I$ is the inter-band hybridization.
We assume a $d$-dimensional hyper-cubic lattice of unit cells containing a $c$- (broad band) and an $f$-orbital each. 
The hopping, of amplitude $t$, is only between nearest-neighbor sites. This gives the dispersion 
$
 \epsilon_{k} =  -2 t \gamma_{\bf k},
$
with $\gamma_{\bf k}=\sum_{\nu=1,d} \cos{k_{\nu}}$.
Unless stated otherwise, we will assume that the inter-band hybridization amplitude is non-zero only 
between $c$ and $f$ orbitals that belong to nearest-neighbor unit cells: $V_{jl}= V_0 \delta_{|{\bf r}_l -{\bf r}_j|,a}$ ($a$ is the lattice parameter). 
In momentum space we have,
\begin{equation}
 V_{\bf k} =  2 V_0 \gamma_{\bf k}~,  ~V_{{\bf k}j} =  e^{-i {\bf k} \cdot {\bf r}_j} \frac{V_{\bf k}}{\sqrt{N}},
 \label{hyb}
\end{equation}
where $N$ is the number of $f$-orbitals and $V_{{\bf k}j}$  is the hybridization amplitude for the
term $c^{\dagger}_{{\bf k}\sigma} f^{}_{j\sigma}$  
($c^{\dagger}_{{\bf k}\sigma} = \frac{1}{\sqrt{N}} \sum_{j} e^{i {\bf k} \cdot {\bf r}_j}  c^{\dagger}_{j \sigma}$).

{\it We will also assume that $\epsilon_f=0$ unless stated otherwise.} In this case, $V_{\bf k}$ cancels at the band crossing points, and for $U \gg 2d\,t$, the Schrieffer-Wolff transformation \cite{Schrieffer66} becomes an expansion in powers of the small parameter $|V_{\bf k} /(\epsilon_{\bf k} -  \epsilon_f) |=|V_0| / |t|$. The resulting effective Hamiltonian is (the detailed derivation will be presented elsewhere \cite{Yifeng10}):
\begin{eqnarray}
H_{\rm eff} &=&\sum_{{\bf k}\sigma} ({\tilde \epsilon}_{\bf k}-\mu) c^\dagger_{{\bf k}\sigma}c^{\;}_{{\bf k}\sigma}+
\sum_{{\bf k} \sigma}  ({\tilde \epsilon}_{f{\bf k}} - \mu) {\tilde f}^\dagger_{{\bf k} \sigma} {\tilde f}^{\;}_{{\bf k}\sigma}
\nonumber \\
&+& J \sum_{\langle j , l \rangle}  ({\bf S}_{j}\cdot {\bf S}_{l} - \frac{1}{4} n^f_{j} n^f_{l}) 
\nonumber \\
&-& \frac{V_0^2}{t} \sum_{\langle j,l \rangle} 
({\bf S}_{j}\cdot {{\bf s}}_{jl} - \frac{1} {4} n^f_j {\hat t}_{jl} ),
\label{heff2}
\end{eqnarray}
where 
${\bf S}_j=\frac{1}{2}\sum_{ss^\prime} {\tilde f}^\dagger_{js}  {\boldsymbol \sigma}_{ss^\prime}  {\tilde f}_{js^\prime}$ is the $f$-electron spin on site $j$,  
$
{\bf s}_{jl} = \frac{1}{2}\sum_{ss^\prime} (c^\dagger_{js} {\boldsymbol \sigma}_{ss^\prime} c_{l s^\prime}
+c^\dagger_{ls} {\boldsymbol \sigma}_{ss^\prime} c_{j s^\prime})
$ is the conduction electron ``bond" spin
(${\boldsymbol \sigma}$ are the Pauli Matrices), 
and 
$
{\hat t}_{jl} =  \sum_{s} (c^{\dagger}_{js} c^{\;}_{ls} + c^{\dagger}_{ls} c^{\;}_{js}).
$
The constrained operators ${\tilde f}^{\dagger}_{j \sigma}= f^{\dagger}_{j \sigma} (1-n^f_{j {\bar \sigma}})$ do not allow for double occupancy of the $f$-orbitals.  
${\tilde f}^{\dagger}_{{\bf k} \sigma}$ creates an electron in the $f$-band with well defined momentum ${\bf k}$: ${\tilde f}^{\dagger}_{{\bf k} \sigma} = \frac{1}{\sqrt{N}} \sum_{j} e^{i {\bf k} \cdot {\bf r}_j} {\tilde f}^{\dagger}_{j \sigma}$.

The original broad band dispersion, $\epsilon_{\bf k}$, is renormalized to 
$
{\tilde \epsilon}_{\bf k}=\epsilon_{\bf k} + \frac{V^2_{\bf k}}{\epsilon_{\bf k}} = \epsilon_{\bf k} - 2 \frac{V^2_0}{t} \gamma_{\bf k}$. The $f$-electrons acquire an effective dispersion
$
{\tilde\epsilon}_{f \bf k}=\epsilon_f - 2 {\tilde t}_f \gamma_{\bf k},
$
where 
$
{\tilde{t}_f} = - V_0^2/t
$
is the effective hopping between nearest-neighbor $f$-orbitals (here we neglected terms ${\cal O}(V_0^2/U)$). The super-exchange interaction $J= 4 {\tilde t}_f^2/U$ is induced by the hopping ${\tilde t}_f $.

Our $H_{\rm eff}$ is an extension of the standard KLM, which is also obtained by applying the Schrieffer-Wolf transformation to the PAM \cite{Schrieffer66}. The important difference between 
that well-known derivation and the one presented here is our original assumption of suppressed hybridization at the band crossing, which leads to a
{\it small control parameter}  for the perturbative expansion. The usual derivation of the KLM \cite{Schrieffer66} 
assumes that the bare $\epsilon_f$ is below the bottom and $\epsilon_f+ U$ is above the top of the broad band to guarantee that the expansion parameter 
$\max \left[|V_{\bf k} /(\epsilon_{\bf k} -  \epsilon_f) |, |V_{\bf k} /( \epsilon_f + U - \epsilon_{\bf k} ) |\right] \ll 1$. This assumption immediately implies 
that the $f$-electrons are localized and, by construction, far from the mixed-valence regime. 
We point out, however, that under certain conditions this assumption is unnecessarily restrictive.  In the case we explicitly consider,  the ratio  $|V_{\bf k} /(\epsilon_{\bf k} -  \epsilon_f) |$, which controls validity of expansion, remains small as long as $|V_0| \ll |t| $. Consequently, $H_{\rm eff}$ remains valid all the way from the localized to the mixed-valence regime.

The qualitative picture introduced by Doniach \cite{Doniach77} for the KLM argues that the $f$-moments 
will order antiferromagnetically if the Ruderman-Kittel-Kasuya-Yosida (RKKY) exchange interaction between local moments is bigger than the Kondo temperature $k_B T_K$. For local Kondo interaction, $J_K$, the  RKKY exchange can be obtained by perturbatively integrating out the itinerant electrons, $J_{RKKY}(|{\bf r}_j - {\bf r}_l|) = J_K^2\int{d{\bf k} \,e^{i{{\bf k} ({\bf r}_j - {\bf r}_l)}} \chi({\bf k}) }$ \cite{RKKY}, with the itinerant static spin susceptibility $\chi({\bf k}) = \sum_{\bf q}{[n_F(\epsilon_{\bf k + q}) - n_F(\epsilon_{\bf q})]/(\epsilon_{\bf k + q} - \epsilon_{\bf q}})$. In particular, for a half-filled band on a hypercubic lattice, the susceptibility diverges at the AFM wave-vector, strongly favoring AFM ordering of the local moments. 
In a similar way, the $c$-degrees of freedom can 
be integrated out in $H_{\rm eff}$, Eq.~(\ref{heff2}); however, the result is different due to the non-trivial momentum space structure of the Kondo interaction in  $H_{\rm eff}$. It is easy to show, that in this case the spin susceptibility in the expression for RKKY exchange has to be replaced by $\tilde \chi({\bf k}) = \sum_{\bf q}({\gamma_{\bf k + q}+\gamma_{\bf q})^2[n_F(\epsilon_{\bf k + q}) - n_F(\epsilon_{\bf q})]/(\epsilon_{\bf k + q} - \epsilon_{\bf q}})$. The $\gamma$ phase factors eliminate the divergence at the AFM wave-vector (in fact, $\tilde \chi({\bf k}_{AFM}) = 0$ for half-filled conduction band), which makes the RKKY interaction  effectively short-ranged and peaked at $k=\pi/2$. The change in the spatial decay power law of the RKKY interaction ($1/r^d \to 1/r^{d+2}$)  results from {\it frustration of the Kondo exchange at the Fermi level}: the usual logarithmic divergence of $\chi({\bf k})$ at $k=2k_F$ is replaced by a logarithmic divergence of  $\partial^2 {\tilde \chi}({\bf k})/{\partial {\bf k}}^2$ at $k=2k_F$. This frustration also suppresses the short range amplitudes of $J_{RKKY}$: $J_{RKKY}(a)\simeq 0.008 V_0^4/t^3$ and $J_{RKKY}(2a)\simeq 0.0186 V_0^4/t^3$.
Fianally, the $f$-electron low-energy sector of $H_{\rm eff}$ is approximately described by a $t-J$ Hamiltonian, $H_{t-J}$, in which the  exchange interaction has two contributions: the AFM super-exchange $J\sim V_0^4/Ut^2$ and a short range $J_{RKKY}\sim V_0^4/t^3$. 
Because of the small numerical prefactor in $J_{RKKY}$, it remains smaller in magnitude than $J$ for $U>2dt$.

The description of $f$ electrons in terms of an effective $t-J$ model has many important consequences. First, it implies that $f$-electron delocalization is induced by doping. 
For $\mu >0$, there is one $f$-electron localized on each orbital and the corresponding moments interact via exchange. This means that the $f$-electrons behave like a MI in the strong 
$U$ limit.  For $\mu \lesssim 0$, a fraction of $f$-electron density is transferred to
the broad band. Although the kintetic energy of the $f$-holes competes against the magnetic ordering, the AFM correlations must survive for a small enough 
concentration of $f$-holes. The simple reason is that the kinetic energy per site scales like $\delta$ (concentration of $f$-holes)
while the magnetic energy per site is proportional to (1-$\delta$). This leads to a phase 
in which the $f$-electrons are simultaneously delocalized and magnetically ordered, with an  
ordered moment comparable to the full moment. The Fermi surface of this phase 
includes small $f$-character hole-pockets that are disconnected  from the big Fermi surface of the broad band.


In the following we present numerical results  computed with the original and the effective 
Hamiltonians in $d=1$. We use the density-matrix renormalization group (DMRG) method \cite{DMRG} to obtain the ground state properties of both Hamiltonians in chains
of $L=20$ unit cells. These calculations have a double purpose. First, we show that the low-energy spectrum of the PAM has two qualitatively different regimes in the mixed-valence state. The first and most traditional ``Kondo-like'' regime takes place when the average hybridization amplitude over the Fermi surface is much stronger than the effective hopping between $f$-orbitals: $|{\bar V}_{k_F}| \gg |{\tilde t}_f|$. This regime is dominated by coherent inter-band charge 
fluctuations. In the second ``Mott-like'' regime, the low-energy physics of the PAM is controlled by coherent intra-band charge fluctuations, i.e.,
the $f$-electrons are well described by a single-band model as it can be inferred from our derivation of $H_{\rm eff}$.
This regime can be stabilized for $|{\bar V}_{k_F}| < |{\tilde t}_f|$. The other purpose is to verify that $H_{\rm eff}$  provides an accurate description of the low-energy spectrum of $H$ as long as $V_{\bf k}$ vanishes at the crossing points.

\begin{figure}[!htb]
\vspace*{-0.5cm}
\hspace*{-0.5cm}
\includegraphics[angle=0,width=8.0cm]{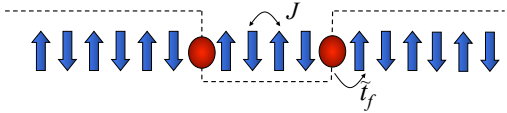}
\vspace{-4.5cm}
\caption{Each hole carries an anti-phase domain wall for the antiferromagnetic correlations 
when a one dimensional Mott insulator is doped away from half-filling.}
\label{antiphase}
\end{figure}

There are several qualitative differences between the Kondo and Mott-like mixed-valence regimes. For $d=1$,
one of these differences appears  in the momentum dependence of the $f$-magnetic structure factor 
$S(q) = \frac{1}{L} \sum_{j,l} e^{i(j-l)q} \langle{\bf S}_j \cdot {\bf S}_l\rangle.$
The Kondo-like mixed-valence regime contains short-range antiferromagnetic fluctuations that lead to a wide peak around $q=\pi$. In contrast, in the Mott-like  mixed-valence state, if the nearest-neighbor AFM interaction dominates ($|J| > |J_{RKKY}|$), the $f$-holes carry an AFM anti-phase boundary \cite{Batista00}. This phenomenon is a direct consequence of the the intra-band nature of the $f$-charge fluctuations: by carrying an anti-phase 
boundary the $f$-holes preserve the antiferromagnetic alignment of nearest-neighbor moments when they hop between different $f$ orbitals (see Fig.\ref{antiphase}).
Consequently, the peak in $S(q)$ shifts to $\pi (1 \pm \delta)$, where $\delta=1-n_f$. 

\begin{figure}[!htb]
\vspace*{-0.42cm}
\hspace*{-0.5cm}
\includegraphics[angle=0,width=7.cm]{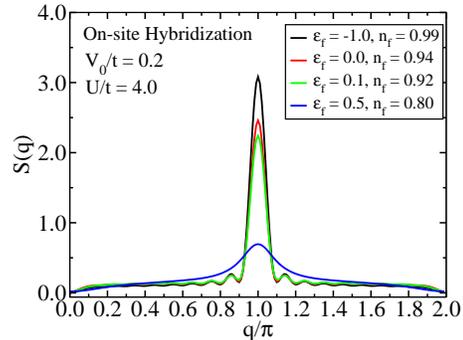}
\vspace{-0.7cm}
\caption{ Magnetic structure factor $S(q)$ for the PAM, $H$, with
on-site hybridization, $V_{\bf k}=V_0$. The results are shown for different $\epsilon_f$ values, thus different 
different $f$-electron densities $n_f$.}
\label{onsite}
\end{figure}

We first analyze the Kondo regime $|{\bar V}_{k_F}| \gg |{\tilde t}_f|$ in which the mixed-valence state is characterized by
coherent {\it inter-band} charge fluctuations. For this purpose, we consider a PAM with on-site
hybridization $V_{\bf k} =  V_0$.
Figure \ref{onsite} shows the DMRG results for different values of $\epsilon_f$. $S (q)$ exhibits a rather sharp peak at $q=\pi$ in the localized regime $n_f \simeq 1$
($\epsilon_f=-1$). The transition to the mixed-valence regime takes place around $\epsilon_f =0$, i.e., $n_f$
becomes significantly lower than one for $\epsilon_f > 0$. The results show that 
the maximum $S (q)$ remains at $q=\pi$, but the peak becomes broader in the mixed-valence state.  
While the dominant magnetic fluctuations are still peaked at $q=\pi$, the effect of the coherent {\it inter-band} 
charge fluctuations is simply to reduce the correlation length of the still-dominant antiferromagnetic correlations.

\begin{figure}[!htb]
\vspace*{-0.42cm}
\hspace*{-0.5cm}
\includegraphics[angle=0,width=8.cm]{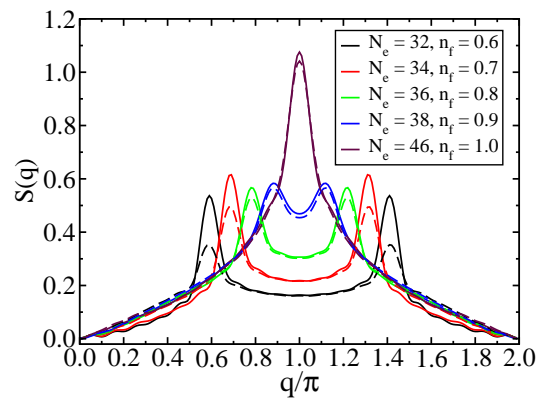}
\vspace{-0.5cm}
\caption{Comparison of $S(q)$ obtained with the PAM and the 
effective Hamiltonian,  $H$ (dashed lines) and $H_{\rm eff}$ (full lines). $V_0/t=0.1$ is the small parameter of the perturbation expansion that leads to $H_{\rm eff}$. Different colours correspond to different values of $N_e$. $L = 20$ unit cells, $U/t = 4.0$.}
\label{eff-PAM}
\end{figure}

The Mott mixed-valence regime can be stabilized under the condition:  $|{\bar V}_{k_F}| \ll |{\tilde t}_f|$. 
To study this regime, we use the hybridization
term introduced in Eq.\eqref{hyb} for which we derived $H_{\rm eff}$ under control. 
Figure \ref{eff-PAM} shows a comparison of the $S(q)$ results computed with $H$ and $H_{\rm eff}$ for 
different values of the total number of electrons $N_e$. Note that $H_{\rm eff}$ reproduces the 
magnetic structure factor obtained with the PAM. This is indeed the expected result because the control parameter 
$V_0/t=0.1$ is small enough to guarantee the validity of our perturbation theory. 
Again, $S(q)$ exhibits a single peak at the AFM wave-vector $q=\pi$ in the localized state ($\mu>0$). 
However, the single peak splits into two symmetric peaks located
at  $q = \pi (1 \pm \delta)$ as the system enters the mixed-valence regime: $\mu \lesssim 0$ and $n_f < 1$. As explained above, this is a clear signature of 
dominant intra-band coherent charge fluctuations. 

\begin{figure}[!htb]
\vspace*{0.2cm}
\hspace*{-0.2cm}
\includegraphics[angle=0,width=9.0cm]{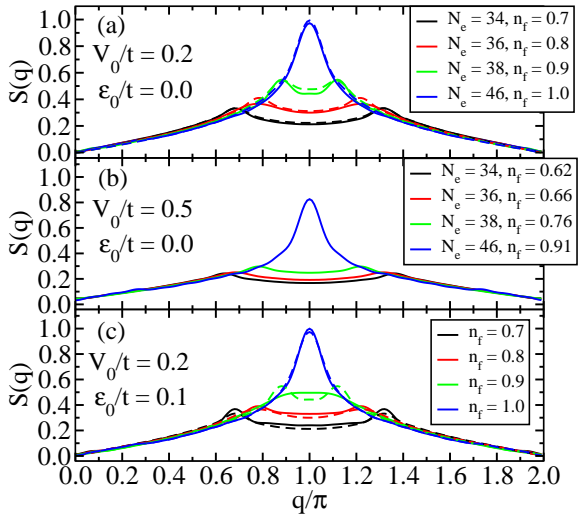}
\vspace{-0.5cm}
\caption{(a) Comparison of $S(q)$ obtained with the PAM, $H$, (full lines) and a pure $t-J$ model for the $f$-electrons (dashed lines) for  
$V_0/t=0.2$ and $\epsilon_f =0$. (b) $S(q)$ for the PAM with $V_0/t = 0.5$ and $\epsilon_f =0$. The double-peak structure remains robust away from the perturbative regime. (c) Same 
comparison as in (a), but for $\epsilon_f=0.1t$ and the same $n_f$ for both models. $L = 20$ units cells, $U/t=4.0$.}
\label{V01}
\end{figure}

To test the relevance of the Kondo exchange term in $H_{\rm eff}$, we also compare the $S(q)$ curve obtained with 
the PAM (for $V_{\bf k}$ given by Eq.(\ref{hyb})) against the results for a pure $t-J$ model, which neglects the Kondo coupling present in Eq. (\ref{heff2}). The comparison is shown
in Fig.~\ref{V01}(a) for $V_0/t=0.2$. The good agreement confirms that $t-J$, as well as the underlying single-band Hubbard model, provides an accurate description of the $f$-electrons in the mixed-valence regime under consideration. As we discuss below, this fact has important implications for more realistic higher-dimensional systems. To test the robustness of the double-peak structure in $S(q)$ away 
from the perturbative regime, we also compute $S(q)$ for the PAM with $V_{\bf k}$ given by Eq.(\ref{hyb}) and $V_0/t=0.5$. 
The results are shown in Fig.~\ref{V01}(b). Although the peaks become broader, the double-peak structure remains robust in the mixed-valence regime. This indicates that the mixed state is still dominated by coherent intra-band fluctuations, 
even away from the perturbative regime $|V_0| \ll |t|$. 

Here we considered a particular form of  $V_{\bf k}$ [Eq.\eqref{hyb}] in order to prove that the mixed valence
regime can be dominated by single-band physics, even starting from  a completely flat bare $f$-band. In general, the hybridization
will not cancel exactly at the Fermi level. 
As an example, in Fig.~\ref{V01}(c) we show a comparison between the $t-J$ and PAM models
for similar values of $n_f$ and $\epsilon_f=0.1t$. This value of $\epsilon_f$ leads to a non-zero hybridization 
at the Fermi level comparable to the $f-f$ effective hopping: $|{\bar V}_{k_F}| \simeq |{\tilde t}_f|/2$. Although the double
peak structure of $S(q)$ gets broadened for $n_f= 0.9$, it remains well defined for $n_f = 0.8$ and $0.7$; this indicates the robustness of the Mott mixed-valence regime proposed here.
Furthermore, a realistic PAM should also include  
a bare $f-f$ hopping $t_{f}$. The single band physics derived in this work remains robust as long as
the average hybridization over the Fermi surface remains smaller than the effective $f-f$ hopping: 
$|{\bar V}_{k_F}| \ll |{\tilde t}_f+t_{f}|$ \cite{Yifeng10}. This observation extends the relevance of our
results beyond the particular PAM considered in this work.  


The physics of lightly doped Mott insulators is well understood in the zero concentration limit (one hole) \cite{Dagotto94}.
As long as the system remains antiferromagnetically ordered, the quasi-particle bandwidth is of order of the effective exchange between local moments. Each quasi-particle consists of a hole ($f$-hole in our case) dressed by a local antiferromagnetic distortion.  The effective mass of the magnetic distortion can be much bigger than the mass of the bare hole, in which case the effective quasi-particle mass $m^*$ is dominated by the exchange interaction,
$m^* \propto 1/J$ (in $d>1$) \cite{Dagotto94}. In this way, heavy fermion behavior can originate from and coexist with local-moment antiferromagnetism (ordered moment comparable to the full moment) in systems with no more than one $f$-electron per ion, such as the Ce-based compounds. 

Our work provides a scenario for the f-electron delocalization that explains several qualitative aspects of a recent dHvA experiment in CeIn$_3$ \cite{Sebastian09}, which cannot be accounted for within the conventional view of the heavy fermion materials. 
In addition, coexistence of local moment AFM and heavy electron superconductivity was observed in the related layered compound CeRhIn$_5$ under pressure \cite{Llobet, Park08}. This is another strong indication that the heavy-fermion behavior can coexist with the local moment AFM.
Future experiments are expected to clarify the applicability of our scenario to these and other heavy fermion compounds.


We thank T. Durakiewicz and J. D. Thompson for useful discussions.
This work was carried out under the auspices of the NNSA of the U.S. Department of Energy at
LANL under Contract No.
DE-AC52-06NA25396 and supported by the LANL/LDRD Program.


\begin{thebibliography}{99}

\bibitem{Sebastian09}
S. E. Sebastian {\it et al.}, Proc. Nat. Acad. Sci. U.S.A. {\bf 106}, 7741 (2009).

\bibitem{Lawrence80}
J. M. Lawrence and S. M.  Shapiro, Phys. Rev. B 22, 4379 (1980).

\bibitem{Doniach77}
S. Doniach, Physica B {\bf 91}, 231 (1977).

\bibitem{Schrieffer66}
J. R. Schrieffer  and P. A. Wolff, Phys. Rev. 149, 491 (1966).

\bibitem{Fisk88}
Z. Fisk {\it et al}, Science {\bf 239}, 33 (1988). 


\bibitem{Trugman90}
S. A. Trugman, Phys. Rev. Lett. {\bf 65}, 500 (1990).

\bibitem{Dagotto94} E. Dagotto, Rev. Mod. Phys. {\bf 66}, 763 (1994).


\bibitem{Yifeng10}
Yi-feng Yang, K. A. Al-Hassanieh,  Ivar Martin and C. D. Batista, in preparation. 

\bibitem{RKKY}
M.A. Ruderman and C. Kittel, Phys. Rev. {\bf 96}, 99 (1954);
T. Kasuya, Prog. Theor. Phys. {\bf 16}, 45 (1956); 
K. Yosida, Phys. Rev. {\bf 106}, 893 (1957).


\bibitem{DMRG}
S. R. White, Phys. Rev. Lett. {\bf 69}, 2863 (1992); Phys. Rev.
B {\bf 48}, 10 345 (1993); K. Hallberg, Adv. Phys. {\bf 55}, 477
(2006); U. Schollw\"ock, Rev. Mod. Phys. {\bf 77}, 259 (2005);
A. F. Albuquerque {\it et al.}, J. Magn. Magn. Mater. {\bf 310}, 1187
(2007).

\bibitem{Batista00}
C. D. Batista and G. Ortiz, Phys. Rev. Lett. {\bf 85}, 4755 (2000).


\bibitem{Llobet} A. Llobet {\em et al.}, Phys. Rev. B {\bf 69}, 024403 (2004).

\bibitem{Park08} T. Park {\it et al.}, Proc. Nat. Acad. Sci. U.S.A. {\bf 105}, 6825 (2008). 

\end{thebibliography}
\end{document}